\documentclass[aps,prd,preprintnumbers,superscriptaddress,twocolumn,nofootinbib]{revtex4-1}
\usepackage{comment}
\usepackage{graphicx}% Include figure files
\usepackage{dcolumn}% Align table columns on decimal point
\usepackage{bm}% bold math
\usepackage{hyperref}% add hypertext capabilities
\usepackage{graphicx} % Required for inserting images
\usepackage{subfigure}
\usepackage{bm}
\usepackage{amsmath,amssymb}
\usepackage{latexsym}
\usepackage{ulem}
\usepackage{datetime}
\usepackage{scrtime}
\usepackage[dvipsnames, svgnames, x11names]{xcolor}
\usepackage{physics}

\definecolor{darkgreen}{RGB}{50,150,0}
\definecolor{purple}{cmyk}{0.5,0.75,0,0}
\definecolor{darkpurple}{RGB}{128,0,128}
\definecolor{ultramarine}{rgb}{0.07, 0.04, 0.56}
\definecolor{cadmiumgreen}{rgb}{0.0, 0.42, 0.24}
\definecolor{indigo(dye)}{rgb}{0.0, 0.25, 0.42}

\hypersetup{colorlinks=true, linkcolor=blue, urlcolor=blue, citecolor=blue}

\def\be{\begin{equation}}
\def\ee{\end{equation}}
\def\ba{\begin{eqnarray}}
\def\ea{\end{eqnarray}}

\def\nn{\nonumber}

\def\d{\mathrm{d}}

\begin{document}
%\raggedbottom

\title{Does Eternal Inflation Violate the Smeared Null Energy Condition?}

\author{Dong-Hui Yu}
\author{Yong Cai}
\email[Corresponding author:~]{caiyong@zzu.edu.cn}
\affiliation{Institute for Astrophysics, School of Physics, Zhengzhou University, Zhengzhou 450001, China}

\begin{abstract}
The smeared null energy condition (SNEC) imposes a semilocal bound on the negative energy accumulated along null geodesics. In eternal inflation, rare stochastic upward fluctuations of the inflaton locally increase the Hubble parameter, creating an apparent tension with the SNEC. Focusing on a canonical single-field model, we investigate whether this quantum-induced self-reproduction violates the SNEC. Using the Fokker-Planck equation, we demonstrate that the ensemble drift of the Hubble parameter is parametrically bounded by slow-roll parameters and semiclassical suppression. Furthermore, a complementary single-trajectory analysis reveals a strong timescale hierarchy, $N_{\rm SNEC} \gg N_{\rm BR}$. This indicates that even for rare upward stochastic excursions, gravitational backreaction invalidates the background spacetime assumption long before the SNEC bound can be mathematically approached. We conclude that while standard stochastic diffusion drives eternal inflation, it does not inherently lead to SNEC violations within the semiclassical slow-roll regime.
\end{abstract}
%\pacs{98.80.-k, 98.80.Cq, 04.50.Kd}

\maketitle
%\tableofcontents

\section{Introduction}
\label{sec:intro}

Energy conditions are physically motivated pointwise conjectures on the stress-energy tensor $T_{\mu\nu}$ that play crucial roles in general relativity and cosmology (see e.g. \cite{Curiel:2014zba,Kontou:2020bta}). Among them, the null energy condition (NEC), defined by $T_{\mu\nu}k^\mu k^\nu \ge 0$ for any null vector $k^\mu$, is generally regarded as the weakest and thus the most difficult to violate classically (see e.g. \cite{Rubakov:2014jja}). However, quantum field theory allows local violations of the pointwise NEC, motivating the development of quantum energy conditions, such as the smeared null energy condition (SNEC) \cite{Freivogel:2018gxj}, see also \cite{Freivogel:2020hiz,Fliss:2021phs}. This condition imposes a semilocal lower bound on the expectation value of the null-projected $T_{\mu\nu}$ smeared along a null geodesic with a smooth weighting function, thereby constraining the accumulation of negative energy in semiclassical gravity.

The interplay between quantum fluctuations and spacetime geometry makes the primordial universe a natural laboratory for testing quantum energy conditions. Recently, the SNEC has served as a diagnostic tool to constrain the viability of nonsingular early universe scenarios, placing nontrivial bounds on the parameter spaces of cosmological bounce \cite{Moghtaderi:2025cns} and Genesis \cite{Yu:2025wak}. Beyond these alternative scenarios, the inflationary paradigm, particularly eternal inflation (see e.g. \cite{Guth:2000ka,Guth:2007ng,Linde:2007fr,Linde:2015edk} for reviews), provides a natural setting for exploring the implications of the SNEC.
In a fixed non-dynamical de Sitter (dS) background, secular growth from perturbative quantum effects leads to a logarithmic accumulation of negative energy \cite{Woodard:2008yt,Kahya:2009sz}, apparently threatening the SNEC. However, as demonstrated in \cite{Yu:2025wak}, once semiclassical backreaction is consistently taken into account, the resulting geometric response becomes significant long before the SNEC bound is approached, rendering the fixed-background dS approximation self-inconsistent.

Nonetheless, a related question arises in the context of eternal inflation. In this scenario, horizon-crossing quantum fluctuations can occasionally dominate over the classical slow-roll evolution, leading to upward fluctuations of the inflaton field and a corresponding increase in the local Hubble parameter $H$, signaling a localized violation of the classical NEC. Such stochastic jumps are the essential ingredient of the self-reproducing picture of eternal inflation \cite{Linde:1986fd}. Does such a quantum-induced increase in the local expansion rate violate the SNEC?

By adopting a ``passive quantum cosmology'' approach, \cite{Kontou:2020pdx} argued that the upward fluctuations associated with eternal inflation need not involve violations of energy conditions, provided one considers the decoherence-induced selection of localized quantum branches in a fixed dS background with negligible gravitational backreaction. While \cite{Easson:2024uxe} investigated whether ``eternal inflation'' violates the SNEC within a specific classical model asymptoting to an Einstein static universe in the infinite past, its kinematic characterization of eternal inflation differs from the conventional stochastic picture of eternal inflation (see e.g. \cite{Guth:2007ng}). Consequently, the questions addressed there are conceptually distinct from that considered in the present work.

In this work, we examine the compatibility between stochastic eternal inflation and the SNEC within the standard semiclassical slow-roll framework. Using both ensemble and single-trajectory analyses, we show that stochastic upward fluctuations do not lead to a violation of the SNEC in the controlled semiclassical slow-roll regime. The underlying reason is that the stochastic contribution remains parametrically suppressed at the ensemble level, while gravitational backreaction becomes important before the SNEC bound can be approached along individual trajectories. We work in Planck units such that the speed of light $c=1$ and the reduced Planck mass $M_{\rm P}\equiv (8\pi G)^{-1/2}$.

\section{Semiclassical Consistency of NEC Violation in dS Space}
\label{sec:dSbackground}

The SNEC acts as a geometric safeguard in semiclassical gravity, preventing quantum corrections from accumulating unbounded negative energy along null trajectories \cite{Freivogel:2018gxj}. Mathematically, it imposes a fundamental lower bound on the smeared expectation value of the stress-energy tensor:
\begin{equation}
\int \d\lambda \,f^2(\lambda) \langle T_{\mu\nu} k^\mu k^\nu \rangle  \geq -\frac{8\pi B M_{\rm P}^2}{\sigma^2}\,,
\label{eq:SNEC}
\end{equation}
where $k^\mu$ is a null vector with affine parameter $\lambda$, $f(\lambda)$ is a smearing test function compactly supported over a characteristic width $\sigma$, and $B$ is a dimensionless constant, see \cite{Freivogel:2018gxj} for details. A benchmark value of $B$ is $B = 1/(32\pi)$.

However, perturbative quantum fields in curved spacetimes often exhibit secular growth, intuitively threatening the bound of SNEC. A quintessential example is a massless, minimally coupled scalar field with a quartic self-interaction ($\lambda \phi^4$) on a fixed, non-dynamical dS background. As computed in standard literature \cite{Kahya:2009sz,Woodard:2008yt}, the expectation value of the null-projected stress-energy tensor accumulates negative energy logarithmically with the scale factor (where $N = \ln a$ is the $e$-folding number):
\begin{equation}
\langle T_{\mu\nu}k^\mu k^\nu\rangle = (\rho+p) (k^0)^2 \approx \frac{\lambda H^4}{(4\pi)^4} \left(-\frac{4}{3} N \right) (k^0)^2\,,
\label{eq:Tkk}
\end{equation}

If one assumes the dS background to be eternally rigid, this secular growth would eventually diverge negatively, ostensibly breaking the SNEC bound. As shown in the Appendix of Ref.~\cite{Yu:2025wak}, a consistent semiclassical treatment naturally resolves this apparent paradox. The theoretical consistency hinges on a timescale competition: the time required for the accumulated negative energy to reach the SNEC bound ($N_{\text{SNEC}}$) versus the timescale at which the background geometry significantly deviates from the exact dS background due to gravitational backreaction ($N_{\text{BR}}$). By choosing the physical smearing time to be of the order of the curvature scale ($\Delta t \sim H^{-1}$), the SNEC integral condition translates to a local density bound $|(\rho+p)_{\rm q}| \lesssim 8\pi B M_{\mathrm{P}}^2 H^2$.\footnote{Throughout the subsequent derivations, the notations ``q'' and ``c'' are adopted to signify contributions from quantum corrections and classical components, respectively.}
By setting the quantum contribution $|(\rho+p)_{\rm q}|\approx \frac{\lambda H^4}{(4\pi)^4}\left(\frac{4}{3}N\right)$ equal to this geometric upper bound ($8\pi B M_{\mathrm{P}}^2 H^2$), we can explicitly solve for $N_{\text{SNEC}}$.

Conversely, to define the backreaction timescale $N_{\text{BR}}$, we consider the energy density dominance. The leading-order quantum energy density grows quadratically with $N$, given by $\rho_{\rm q} \approx \frac{\lambda H^4}{(4\pi)^4} 2 N^2$. We define $N_{\text{BR}}$ as the crucial moment when this secularly growing quantum energy density reaches a noticeable fraction $\varepsilon$ of the background energy density $3M_{\mathrm{P}}^2H^2$. As discussed in Ref.~\cite{Yu:2025wak}, comparing these two timescales yields a pronounced hierarchy:
\begin{equation}
\frac{N_{\text{SNEC}}}{N_{\text{BR}}} \approx \pi B \sqrt{\frac{72}{\varepsilon \lambda}} \left( \frac{M_{\mathrm{P}}}{H} \right) \gg 1 \,,
\end{equation}
Since the scale ratio $M_{\mathrm{P}}/H$ is typically very large, e.g., $\mathcal{O}(10^5)$ in the inflationary universe, and $\rho$ grows as $N^2$ while $|(\rho+p)_{\rm q}|$ grows only as $N$, this ratio is overwhelmingly greater than unity.

This hierarchy ($N_{\rm SNEC}\gg N_{\rm BR}$) indicates that the backreaction will destroy the de
Sitter background long before the SNEC bound is challenged, see the Appendix of Ref.~\cite{Yu:2025wak} for more details. In this sense, the SNEC is protected within the regime where the semiclassical calculation is reliable.

\section{SNEC Constraints on Stochastic Eternal Inflation}
\label{sec:eternal}

In this section, we transition our focus from a fixed dS background to the highly dynamic regime of eternal inflation. Utilizing the stochastic dynamics framework (see e.g. \cite{Starobinsky:1986fx,Starobinsky:1994bd}), we show that stochastic eternal inflation is consistent with the SNEC within the semiclassical slow-roll regime.

For simplicity, we work with a canonical single-field inflation model. The action is given by
\begin{equation}S=\int\mathrm{d}^4x\sqrt{-g}\left[\frac{M_{\rm P}^2}{2}R-\frac{1}{2}\partial_\mu\phi\partial^\mu\phi-V(\phi)\right]\,,\label{eq:action260605}
\end{equation}
where the scalar field $\phi$ is the inflaton.
The background evolution is governed by the equations of motion
\begin{align}
    &\ddot{\phi}+3H\dot{\phi}+V'(\phi) = 0\,,\\
    &3M_{\rm P}^2 H^2 =\frac{1}{2}\dot{\phi}^2+ V(\phi)\,,
\end{align}
with dots denoting derivatives with respect to cosmic time $t$ and primes derivatives with respect to $\phi$. In the slow-roll regime, the classical field drift over a Hubble time $\Delta t \sim H^{-1}$ is $\Delta \phi_{\rm c} \approx -V'/(3H^2)$. Concurrently, horizon-crossing quantum fluctuations induce stochastic jumps with a characteristic amplitude $\Delta \phi_{\rm q} \simeq H/(2\pi)$ \cite{Starobinsky:1982ee,Vilenkin:1982wt,Linde:1982uu}. When the stochastic diffusion locally opposes and dominates the classical drift ($|\Delta \phi_{\rm q}| > |\Delta \phi_{\rm c}|$), the field in certain rare, ``lucky" Hubble patches increases, leading to an increase of the local Hubble parameter. This phenomenon is known as eternal inflation, for a detailed discussion, see e.g., \cite{Linde:1986fd,Starobinsky:1986fx,Starobinsky:1994bd}.

To describe this within the stochastic framework, the eternally inflating patch is treated as an open system continuously receiving short-wavelength quantum modes crossing the dS horizon. The semiclassical Einstein equation $G_{\mu\nu} = M_{\rm P}^{-2} T_{\mu\nu}^{\text{eff}}$ is driven by an effective stress-energy tensor that can be conceptually decomposed as $T_{\mu\nu}^{\text{eff}} = T_{\mu\nu}^{(\text{c})} + T_{\mu\nu}^{(\text{q})}$.
For the pure classical component of the Lagrangian density given by Eq. (\ref{eq:action260605}),
the projection of $T_{\mu\nu}^{(\text{c})}$ along a null vector $k^\mu$ strictly satisfies the NEC, i.e., $T_{\mu\nu}^{(\text{c})} k^\mu k^\nu = (\partial_\mu\phi k^\mu)^2 \ge 0$.
The increase of the local Hubble parameter $H$ is sourced by the stochastic quantum component $T_{\mu\nu}^{(\text{q})}$, which violates the classical NEC, i.e., $T_{\mu\nu}^{(\rm q)}k^\mu k^\nu
=(\rho_{\rm q}+p_{\rm q})(u_\mu k^\mu)^2<0$.

The SNEC bounds the semiclassical expectation value $\langle T_{\mu\nu}^{\text{eff}} k^\mu k^\nu\rangle$ integrated along a null geodesic, i.e., $\int \d\lambda\, f^2(\lambda)\langle T_{\mu\nu}^{\rm eff}k^\mu k^\nu\rangle$, where $f^2(\lambda)$ is the smearing function.
Does the quantum-induced increase of the local Hubble parameter necessarily lead to a violation of the SNEC?

\subsection{Ensemble-Level Constraints}
\label{subsec:ensemble}

In fact, the ensemble drift is controlled by the competition between classical drift and quantum diffusion. This coarse-grained dynamics is governed by the Fokker-Planck equation (see eg. \cite{Starobinsky:1986fx,Starobinsky:1994bd,Linde:1986fd}) for the normalized probability distribution $P(\phi,t)$:
\begin{equation}
\frac{\partial P}{\partial t} = \frac{\partial}{\partial\phi}\left(\frac{V'P}{3H}\right) + \frac{1}{2}\frac{\partial^2 }{\partial \phi^2}\left(\frac{H^3}{4\pi^2}P\right).
\end{equation}
Here, following the distinction between coordinate-volume and physical-volume distributions in stochastic eternal inflation~\cite{Linde:1986fd}, we use a probability distribution over comoving Hubble patches.
It is normalized over the inflaton field,
\begin{equation}
\int \d\phi\,P(\phi,t)=1.
\end{equation}
This choice still includes stochastic diffusion and rare upward fluctuations, but does not include the additional physical-volume weighting of faster-expanding patches.
Since the SNEC is a local semiclassical bound along a null geodesic, this is the ensemble relevant for the local SNEC estimate considered here.

For a function $F(\phi)$, its ensemble expectation value at a given time is defined as
\begin{equation}
\ev{F(\phi)}=\int F(\phi)P(\phi,t)\d\phi\,.
\end{equation}
The time evolution of this expectation value can then be written as
\begin{equation}
\frac{\d}{\d t}\ev{F(\phi)}=\int F(\phi)\frac{\partial P(\phi,t)}{\partial t}\d\phi\,.
\end{equation}
Substituting the evolution equation for $P(\phi,t)$ and performing integration by parts, we assume that the probability distribution and the probability flux vanish sufficiently fast at the boundary, or that boundary conditions are chosen such that the boundary terms do not contribute. One obtains
\begin{equation}
\frac{\d}{\d t}\ev{F}=\ev{-\frac{V'}{3H}F'+\frac{H^3}{8\pi^2}F''}.
\label{Eq:average}
\end{equation}

We have $\ev{T_{\mu\nu}^{\text{(eff)}}k^{\mu}k^{\nu}}\simeq-2\dot{H}_{\rm eff}M_{\rm P}^2(k^0)^2$. In the stochastic coarse-grained description, the single-realization $\dot{H}$ is not a classical smooth quantity. The quantity entering the effective SNEC estimate is the ensemble drift of the Hubble parameter, which we denote by
$\dot H_{\text{eff}}\equiv \d\langle H\rangle/\d t$.

In the slow-roll regime, the local Hubble parameter $H$ in a patch can be regarded as an explicit function of $\phi$, with $3M_{\rm P}^2H^2(\phi)\simeq V(\phi)$. Taking $F=H(\phi)$ in Eq.~\eqref{Eq:average}, we obtain
\begin{equation}
\frac{\d}{\d t}\ev{H}
=\ev{-\frac{V'}{3H}H'+\frac{H^3}{8\pi^2}H''}.
\end{equation}
Using the slow-roll parameters $\epsilon=M_{\rm P}^2(V'/V)^2/2$ and $\eta=M_{\rm P}^2V''/V$, this becomes
\begin{equation}
\frac{\d}{\d t}\ev{H}=-\ev{\epsilon H^2}+\frac{1}{16\pi^2M_{\rm P}^2}\ev{H^4(\eta-\epsilon)} .
\label{Eq:H-drift}
\end{equation}
It is worth noting that although eternal inflation is driven by stochastic upward jumps of $\phi$, the jump amplitude $\Delta \phi_{\rm q} \simeq H/(2\pi)$ is parametrically small compared to the Planck scale ($H \ll M_{\mathrm{P}}$). In the flat regions of the potential where eternal inflation occurs, such jumps induce negligible relative changes in $V(\phi)$ and its derivatives. Consequently, the slow-roll parameters $\epsilon$ and $\eta$ remain $\ll 1$ within these ``lucky'' patches, ensuring the validity of the slow-roll approximation.

Substituting Eq.~\eqref{Eq:H-drift} into Eq.~\eqref{eq:SNEC} gives the condition for SNEC preserving the bound,
\ba
&\,&\int \d\lambda\, f^2(\lambda)(k^0)^2\Big[-\ev{\epsilon H^2}
\nn\\
&\,&\qquad\qquad +\frac{1}{16\pi^2M_{\rm P}^2}\ev{H^4(\eta-\epsilon)}\Big]\le\frac{4\pi B}{\sigma^2}.
\ea
The smearing interval is chosen to be one stochastic coarse-graining time at the ensemble level, so that $\sigma^{-2}\simeq (k^0)^2\ev{H^2}$. Over this short interval we evaluate the Fokker-Planck ensemble at the central time $t_*$. The patch dependence of $H(\phi)$ is not discarded. Therefore, using the normalization of $f(\lambda)$, $\int f^2(\lambda)\d \lambda=1$, the SNEC reduces to
\begin{equation}
-\ev{\epsilon H^2}+\frac{1}{16\pi^2M_{\rm P}^2}\ev{H^4(\eta-\epsilon)}\le4\pi B\ev{H^2}.
\label{Eq:ensemble-SNEC}
\end{equation}
Since $\epsilon\ge0$, the classical drift term is non-positive, i.e., $-\ev{\epsilon H^2}\le0$. Since different patches have different local values of $H(\phi)$, we define $H_{\max}$ as the maximal Hubble parameter within the ensemble, and similarly take $\delta_m$ as an upper bound of $|\eta-\epsilon|$. Then
\begin{equation}
\ev{H^4(\eta-\epsilon)}\le\ev{H^4|\eta-\epsilon|}\le
\delta_m\ev{H^4}\le\delta_m H_{\max}^2\ev{H^2}.
\end{equation}
Therefore,
\begin{equation}
\frac{1}{16\pi^2M_{\rm P}^2}\ev{H^4(\eta-\epsilon)}
\le
\frac{\delta_m H_{\max}^2}{16\pi^2M_{\rm P}^2}\ev{H^2}.
\end{equation}

Since the classical drift term in Eq.~\eqref{Eq:ensemble-SNEC} is non-positive, a sufficient condition for the ensemble-averaged SNEC to be satisfied is that the stochastic-diffusion contribution is smaller than the SNEC scale.
As a result, we can rewrite the SNEC condition as
\begin{equation}
\mathcal A_{\rm SNEC}\equiv\frac{\frac{1}{16\pi^2M_{\rm P}^2}\ev{H^4(\eta-\epsilon)}}{4\pi B\ev{H^2}}\le 1.
\end{equation}

Using the bound above, one obtains
\begin{equation}
\mathcal A_{\rm SNEC}\le\frac{\delta_m}{64\pi^3B}\frac{H_{\max}^2}{M_{\rm P}^2}\,.\label{eq:26060512}
\end{equation}
For the benchmark value $B=1/(32\pi)$, (\ref{eq:26060512}) becomes
\begin{equation}
\mathcal A_{\rm SNEC}\le\frac{\delta_m}{2\pi^2}\frac{H_{\max}^2}{M_{\rm P}^2}.
\end{equation}
In the semiclassical slow-roll regime, we have $\delta_m\ll1$, $H_{\max}^2 \ll M_{\rm P}^2$, and hence
\begin{equation}
\mathcal A_{\rm SNEC}\ll1.
\end{equation}
Thus the ensemble-averaged SNEC is parametrically satisfied: stochastic upward fluctuations are too small to overcome the SNEC bound within the semiclassical slow-roll regime.

\subsection{Single-Trajectory Analysis}
\label{subsec:single}

While the ensemble estimate is parametrically bounded, what if we track a single, exceptionally ``lucky" bubble where the field undergoes an upward stochastic excursion? To investigate this single-patch dynamics, we employ the coarse-grained Langevin equation \cite{Starobinsky:1986fx,Starobinsky:1994bd}:
\begin{equation}
\dot{\phi} = -\frac{V'(\phi)}{3H} + \xi(t,\vec{x}) \,,
\end{equation}
where $\xi(t,\vec{x})$ represents the stochastic force, and $\xi(t)$ gives ($\vec{x}$ is the same throughout)
\begin{equation}
\langle \xi(t_1)\xi(t_2) \rangle = H^3(4\pi^2)^{-1}\delta(t_1-t_2).
\end{equation}
By integrating the stochastic force, the accumulated variance over an $e$-folding interval $\Delta N$ in a dS background exhibits a secular growth:
\begin{equation}
\langle (\delta\phi)^2 \rangle \approx \frac{H^2}{4\pi^2}\Delta N
\label{eq:deltaphi}
\end{equation}

Following the Appendix of Ref.~\cite{Yu:2025wak} and Sec.~\ref{sec:dSbackground}, we first calculate the characteristic timescale $N_{\text{SNEC}}$ required to mathematically approach the SNEC bound. Approximating the SNEC integral locally over $\Delta t\sim H^{-1}$, with $\Delta\lambda\approx(k^0H)^{-1}$, SNEC gives a bound of order $\dot H\lesssim4\pi B H^2$. However, along a single Langevin trajectory, the instantaneous $\dot H$ is not a well-defined smooth quantity because of the stochastic kicks.
We therefore use the root-mean-square (RMS) size of the accumulated stochastic displacement to estimate how much the quantity $\dot H(\phi)\simeq -V'^2/(6V)$ is shifted. The RMS fluctuation is given by a first-order Taylor expansion $\Delta \dot{H} \approx \left| \frac{\d\dot{H}}{\d\phi} \right| \sqrt{\langle (\delta \phi)^2 \rangle}$, where $\frac{\d\dot{H}}{\d\phi} = -\frac{V'}{3M_{\rm P}^2}(\eta-\epsilon)$. Equating this RMS estimate $\Delta\dot H$ to $4\pi B H^2$, i.e.,
\be
\frac{|V'||(\eta-\epsilon)|}{3M_{\rm P}^2}\frac{H}{2\pi}\sqrt{N_{SNEC}}\approx 4\pi BH^2 \,,
\ee
we find
\be
 N_{\text{SNEC}} \approx \left( \frac{24\pi^2 B H M_{\rm P}^2}{|\eta - \epsilon| |V'|} \right)^2 .
\ee

Next, we evaluate the backreaction timescale, $N_{\text{BR}}$, for a single stochastic trajectory. The RMS backreaction timescale is defined as the number of $e$-folds at which the RMS fluctuation of the energy density reaches a critical fraction $\varepsilon$ of the background density. In the slow-roll regime, the total energy density is predominantly potential-dominated, i.e., $\rho \approx V(\phi)$. For a stochastic displacement $\delta\phi$ induced by horizon-crossing modes, the resulting energy density fluctuation can be approximated to first order as $\delta\rho \approx V'(\phi) \delta\phi$. The corresponding RMS fluctuation is then given by
\begin{equation}
 \Delta \rho = \sqrt{\langle (\delta \rho)^2 \rangle} \approx |V'(\phi)| \sqrt{\langle (\delta \phi)^2 \rangle} \,.
 \end{equation}
Using \eqref{eq:deltaphi}, we find the RMS energy density growth as a function of $e$-folds:
\begin{equation}
 \Delta \rho \approx |V'(\phi)| \frac{H}{2\pi} \sqrt{\Delta N}\,,
\end{equation}
The semiclassical background is assumed to be stable as long as the fluctuations remain a small fraction of the total density, i.e., $\Delta \rho \lesssim \varepsilon \rho_{\text{BG}}$, where $\rho_{\text{BG}} \approx 3 M_{\rm P}^2 H^2$. Defining $N_{\text{BR}}$ as the point where this equality is saturated, i.e.,
\begin{equation}
 |V'|\left(\frac{H}{2\pi}\sqrt{N_{BR}}\right)= \varepsilon 3M_{\rm P}^2H^2
\end{equation}
we obtain
\be
N_{\text{BR}} \approx \left( \frac{6\pi M_{\rm P}^2 \varepsilon H}{|V'|} \right)^2 \,.
\ee

Comparing the two timescales, we find
\begin{equation}
 \frac{N_{\text{SNEC}}}{N_{\text{BR}}} = \left( \frac{4\pi B}{|\eta - \epsilon| \varepsilon} \right)^2\,.
 \label{Eq:NsencNbr}
\end{equation}
Since $|\eta-\epsilon|$ is suppressed in the slow-roll regime, this ratio is pretty large.
For representative slow-roll values comparable to those inferred on observational scales,
$\epsilon\lesssim\mathcal O(10^{-3})$ and $|\eta|\sim\mathcal O(10^{-2})$ \cite{BICEP:2021xfz,Planck:2018jri}, one obtains, for example,
\begin{equation}
\frac{N_{\rm SNEC}}{N_{\rm BR}}\gtrsim \mathcal O(10^4)\gg1
\end{equation}
for $\varepsilon\simeq {\cal O}(0.1)$ and $B=1/(32\pi)$. This hierarchy indicates that even for a ``lucky'' upward stochastic excursion, the RMS backreaction scale is reached long before the accumulated fluctuation can approach the SNEC bound.

\section{Conclusion and Discussion}\label{Sec:con}

In this work, we investigated whether stochastic eternal inflation violates the SNEC. Focusing on a canonical single-field inflation model within the semiclassical slow-roll regime, we demonstrated that the SNEC is parametrically preserved. At the ensemble level, the normalized Fokker-Planck dynamics keep the upward stochastic drift strictly bounded by the slow-roll parameters, specifically $|\eta-\epsilon|$, and the semiclassical suppression factor $H_{\text{max}}^2/M_{\rm P}^2 \ll 1$. Complementary to this, our single-trajectory analysis confirms that even for rare upward stochastic excursions, gravitational backreaction invalidates the background spacetime assumption long before the accumulated fluctuations can approach the SNEC bound, yielding a strong hierarchy $N_{\rm SNEC} \gg N_{\rm BR}$.

This conclusion relies on the validity of the slow-roll and semiclassical approximations. Consequently, it applies naturally to standard scenarios like chaotic inflation and new (hilltop) inflation (see e.g. \cite{Guth:2000ka,Guth:2007ng,Linde:2007fr,Linde:2015edk}). For instance, near the hilltop, the limits $\epsilon \to 0$ and $|\eta| \ll 1$ inherently satisfy the required bounds $|\eta-\epsilon| \ll 1$ and $H_{\text{max}}^2/M_{\rm P}^2 \ll 1$, making a separate potential-specific analysis unnecessary. However, it is important to explicitly state the limitations of our findings. Because our framework is based on a canonical single-scalar field, we cannot definitively conclude that the SNEC is universally protected in all variations of eternal inflation. In more complex scenarios, such as non-canonical multi-field inflation models, or regimes deeply in the quantum gravity domain where the semiclassical treatment breaks down, different dynamics might emerge. Therefore, our results specifically answer that while standard slow-roll stochastic diffusion drives self-reproduction, it does not lead to SNEC violations within the regime where semiclassical gravity remains valid.

Finally, we note that the stochastic NEC violation discussed here is phenomenologically distinct from the classical NEC-violating phases often studied in effective field theories \cite{Cai:2016thi,Cai:2017tku,Cai:2017dyi,Cai:2017dxl,Cai:2020qpu}, see also \cite{Creminelli:2016zwa,Nishi:2016ljg,Ilyas:2020qja,Ilyas:2020zcb,Zhu:2021whu,Ye:2025pem}. In those scenarios, used to generate primordial black holes \cite{Cai:2023uhc} and associated gravitational-wave signals \cite{Cai:2020qpu,Cai:2022nqv,Cai:2022lec,Ye:2023tpz,Pan:2024ydt,Jiang:2024woi,Yu:2026vey}, the NEC violation occurs directly at the level of the background evolution. For these systems, the SNEC is expected to act primarily as a constraint on the classical background trajectory itself. Furthermore, the interplay between such a classical NEC-violating background and its superimposed quantum fluctuations could introduce novel stochastic dynamics. A detailed evaluation of how the SNEC restricts these classical scenarios and their associated stochastic behaviors remains an intriguing direction for future work.
\\

\acknowledgments

We thank Mian Zhu for his collaboration on Ref.~\cite{Yu:2025wak}, and we are also grateful to the anonymous referee of that paper, whose insightful comments inspired the present work.
This work is supported in part by the National Natural Science Foundation of China (Grant No. 12575066) and the Natural Science Foundation of Henan Province (Grant No. 262300421236).

\bibliography{necv}
\bibliographystyle{utphys}

\end{document}